\documentclass[aps,prl,twocolumn,superscriptaddress]{revtex4}

\usepackage{amsmath}
\usepackage{graphicx}

\begin{document}

%\preprint{}

\title{Phase transition for cutting-plane approach to vertex-cover problem}


\author{Timo Dewenter}
\author{Alexander K. Hartmann}
\email[]{a.hartmann@uni-oldenburg.de}
\homepage[]{http://www.compphys.uni-oldenburg.de/en/}
\affiliation {Institut f\"ur Physik, Universit\"at Oldenburg, D-26111
Oldenburg, Germany }

\date{\today}

\begin{abstract}
We study the vertex-cover problem which is an NP-hard optimization
problem and a prototypical model exhibiting phase transitions
on random graphs, e.g., Erd\H{o}s-Renyi (ER) random graphs.
 These phase transitions coincide with changes
of the solution space structure, e.g, for the ER ensemble
at connectivity $c=e\approx 2.7183$ from replica symmetric to
replica-symmetry broken. For the vertex-cover problem, also the 
typical complexity of exact branch-and-bound algorithms, which proceed
by exploring the landscape of feasible configurations, change close
to this phase transition  from ``easy'' to ``hard''.
In this work, we consider an algorithm which has a completely different
strategy: The problem is mapped onto a linear programming problem
augmented by a cutting-plane approach, hence the algorithm operates
in a space \emph{outside} the space of feasible configurations until the final
step, where a solution is found. Here we show that this type
of algorithm also exhibits  an  ``easy--hard'' transition
around $c=e$, which strongly indicates that the typical 
hardness of a problem is fundamental to the problem and 
not due to a specific representation of the problem.
\end{abstract}


\maketitle

NP-hard combinatorial optimization problems \cite{garey1979,papadimitriou1998} 
are fundamental to computational complexity, because despite much
effort no algorithm has been found so far, which is able to solve
these problems in the \emph{worst case} in polynomial time,
leading to the famous \emph{P-NP problem}. One way to try to understand the
root of the apparent computational hardness is to analyze hard instances
of problems. This has attracted  much 
interest in statistical physics 
\cite{phase-transitions2005,mezard2009,moore2011}.
Phase transitions on suitably chosen
ensembles of random instances were found, e.g., for the 
Satisfiability Problem \cite{kirkpatrick1994},
  the Traveling Salesman Problem
\cite{gent1996} or the vertex-cover problem (VC)
\cite{cover2000}.
For  exact branch-and-bound
 algorithms \cite{davis1962,sedgewick1990} the hardest instances
are found right at these phase transitions, often related
to a change from a typically polynomially (``easy'')
to a typically exponentially (``hard'') region.
Branch-and-bound algorithms systematically explore the space
of feasible solutions (branching) while trying to avoid uninteresting 
configurations via updating efficient bounds.
The behavior of these exact algorithms can be partially understood in terms
of an effective dynamics inside the phase diagrams 
\cite{cover-time2001,cocco2001}.
In practice very efficient but not exact are stochastic algorithms, 
e.g. WalkSAT \cite{papadimitriou1991} 
or ASAT \cite{ardelius2006} and  message-passing algorithms 
\cite{mezard2002,mezard2002b}, 
inspired by statistical mechanics methods like the cavity approach 
\cite{mezard1987}. Also these types of algorithms rely on either
moving in configuration space or on
 calculating iteratively probabilities (weights) for different subspaces
of configurations.

Here, we consider a completely different and complementary type of algorithm,
\emph{linear programming} (LP), which is a standard approach
for practical optimization problems \cite{papadimitriou1998}.
 In connection with 
\emph{cutting planes} (CP) \cite{cook1998}, it is a very efficient (but 
apparently still
worst-case exponential) approach to combinatorial optimization
problems. This approach is fundamentally different from the algorithms
mentioned above since it does not move inside the configuration space
but instead 
considers non-feasible (non-combinatorial) assignments to the variables
which are always more optimal than the true feasible solution.
Cutting planes are constraints which are added additionally and iteratively
to the problem until a feasible solution is found, which is then
the optimal solution. In particular we study the vertex cover problem
via LP and CP for Erd\H{o}s-Renyi random graphs \cite{erdoes1960}. 
We show that 
VC with our LP/CP implementation changes from ``easy'' to ``hard''
right at the same transition point, where this change occurs for
a branch-and-bound algorithm, and where the solution landscape
changes from simple (replica symmetric in the spin-glass
language \cite{mezard1987}) to complex (replica-symmetry broken).
Hence, our results indicate that the typical hardness of a problem
seems to be quite universal since the changes from ``easy'' to ``hard''
are visible for algorithms which are based on fundamentally different
notions of configuration space.

\paragraph*{Model}
Let $G = (V,E)$ be an undirected graph with  $N$ vertices $i\in V$ 
and $M$ edges $\{i,j\}\in E$. A vertex cover $V_{\text{VC}} \subset V$ is a 
subset of vertices so that for all edges $\{i,j\} \in E$ 
at least one end $i$ or $j$  is contained in
$V_{\text{VC}}$. The vertices $i\in V_{\text{VC}}$ are
called \emph{covered}, \emph{uncovered} else.
 We are interested in vertex covers of $G$ of minimum cardinality
$|V_{\text{VC}}|$, the \emph{minimum vertex covers}.
 The decision problem if a VC with fixed cardinality
 exists or not belongs to the class of NP-complete problems \cite{garey1979}.

The analytical solution of VC on Erd\H{o}s-Renyi graphs  
exhibits a phase transition  at the average connectivity $c=e\approx 2.7183$:
for $c<e$, the solution is replica symmetric, while for $c>e$
replica symmetry breaking was found \cite{cover2000}. This can be
seen also numerically when clustering the minimum
vertex covers \cite{vccluster2004}. Furthermore,
in connection with the \emph{leaf-removal heuristic} \cite{bauer2001}, 
the typical-case
complexity of a branch-and-bound algorithm changes form ``easy'' to ``hard''
at $c=e$.

\paragraph*{Linear-Programming Approach}
First, we translate the VC problem to an integer 
linear programming (ILP) problem  \cite{papadimitriou1998}, each
of the $N$ nodes of the graph is represented by a variable $x_i \in
\{0,1\}, \; i=1,\ldots,N$. The value $x_i = 1$ denotes a covered, $x_i = 0$
indicates an uncovered  node. The fact that for each edge $\{i,j\}$
$i$ or $j$ must be covered can be written as
$x_i + x_j \geq 1$. Minimizing the cardinality of the cover
means we want to minimize $\sum_i x_i$.
When we relax the integer constraint to
$x_i \in [0,1]$, the set of constraints $x_i + x_j \geq 1$ describes
a \emph{polytope}. Now  we obtain the following  linear programming 
problem (LP):
\begin{tabbing}[tb]
 \hspace*{1.7cm}\=\kill
 \textbf{Minimize}    \>$x = \sum_{i=1}^N x_i$ \\[2mm]
 \textbf{Subject to}  \>$0\le x_i \le 1 \; \forall \; i \in V $\\
                      \>$x_i + x_j \geq 1 \quad \forall \; \{i,j\} \in E$ \\ 
\end{tabbing}
This can be solved efficiently, i.e., typically in polynomial time,
by the simplex algorithm (SX)
\cite{papadimitriou1998,dantzig1948}. 
We used the public available {\tt lp\_solve} \cite{lp_solve}
 with Bland's \emph{first index pivoting}
\citep{bland1977}. Note that now the solutions
are not guaranteed to be integer-valued any more, 
variables with $x_i \in \; ]0,1[$ we call \emph{undecided}. 
Such solutions we call \emph{incomplete}. 
The value of $x=\sum_i x_i$
is always a lower bound for the cardinality of a complete solution.
On the other hand,
if a solution computed by SX 
is \emph{complete}, i.e., all variables are integer-valued,
then it is immediately clear that it is a correct
minimum for VC.

\paragraph*{Cutting-Plane Approach}

In order to obtain more complete solutions, the CP approach \cite{cook1998}
can be used. The basic idea is to limit the solution space by adding
extra constraints, which exclude incomplete solutions. In principle,
many types of extra constraints are possible. Here, we apply the following
heuristics, inspired by the nature of the problem: For any graph which is
a cycle, setting all variables of non-isolated nodes 
to $0.5$ (0 else) is a solution of the LP, 
but incomplete.
Nevertheless, for a cycle of odd length $l=2k+1$, the size of the minimum
cover is always $k+1>k+0.5$. Hence, for any cycle of length $l=2k+1$ 
in a graph,
at least $k+1$ nodes of the cycle must be covered.
Thus, after an execution of SX, if the solution is incomplete,
we try to detect cycles of odd length $l$ where the condition
\begin{equation}
	\sum_{x_i \in \text{loop}} x_i \geq \left \lceil \frac l 2
	\right \rceil
	\label{constr}
\end{equation} 
is violated ($\lceil l/2 \rceil$ is the largest integer larger or
equal to $l/2$) and add this constraint to the LP. Technically,
 the loops are obtained by searching a random spanning tree
(ST) in the graph via a random breadth-first search and adding a randomly
chosen edge, which is part of the graph but not of the ST.
Our algorithm stops, if $s = 20 M$ times for a randomly chosen spanning trees
and for all loops emerging from these trees
we did not add a constraint (because the loop was of even length
or the constraint was already fulfilled by the current incomplete
solution). Otherwise, SX is executed for the next time and the solution
checked for completeness again.

In general,
SX guarantees to obtain a solution in a corner of the polytope
with minimum $x$. This means that
still non-integer solutions can be obtained,how frequently
depends also on the heuristics used in the actual SX implementation.
Anyway, an incomplete solution obtained by CP+SX 
provides another lower bound $x$, usually
better (but never worse) than that obtained by SX alone.

\paragraph*{Node heuristics}
To complete an incomplete solution,
we also applied the following ``node'' heuristics (NH):
It randomly selects a vertex $i$
with an undecided variable $x_i \in \; ]0,1[$ and adds $x_i = 0$ to the
LP and solves it again. This forces the SX algorithm to set 
nodes $j$ adjacent to $i$ to $x_j = 1$. After each run of the SX
algorithm we checked whether still undecided variables are found and if
necessary the procedure is repeated. This ensures that finally a
complete solution, i.e., a vertex cover is found, but it doesn't have to be a
minimum one. Hence the values of $x$ obtained in this way are
upper bounds to the true minimum vertex covers.
Note that we also tried a heuristics
where $x_j = 1$ is added to the LP, but it
provided typically higher values of $x$ as solution,
in particular for large graph connectivity $c > 5$.

\paragraph*{Results}

Next, simulation results for the different types of
 algorithms are presented for ER random
graphs  of $N$ nodes, up to $N=280$. We used the ensemble
where for each graph $M$ edges are created randomly with uniform weight,
i.e., the connectivity is $c=2M/N$.

\begin{figure}[htb]
   \includegraphics[width=0.47\textwidth]{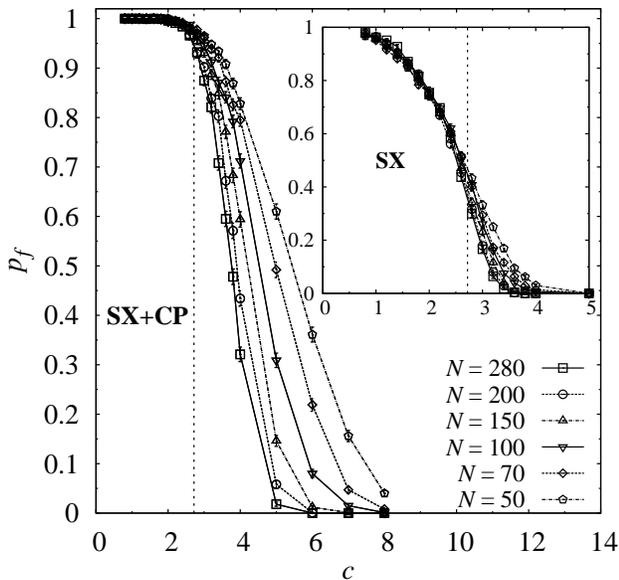}
   \caption{Fraction $p_f$ of complete solutions  for CP approach as a
   function of connectivity $c$ for ER random graphs.
   The inset shows $p_f$
   for the SX algorithm. The vertical line denotes $c=e$,
the other lines are guides to the eyes only.\label{Num_sol}}
 \end{figure}

Fig.\ \ref{Num_sol} shows the fraction  $p_f$ of graphs
which exhibit a complete solution, for the SX (inset) and SX+CP algorithms,
obtained from averaging over 1000 realizations of graphs and for different
system sizes, respectively. Apparently, SX is able to find solutions
up to about $c=1$, where a sharp drop of $p_f$ is visible,
resembling a phase transition. Note that the percolation transition
of the ER ensemble is at $c=1$. For $c<1$, ER random graphs consist
mainly of trees, which are apparently easy to solve, even for SX.
When including CP,  more samples can be solved, the transition
shifts to a point close to the $c=e$, where replica symmetry breaking occurs
and where the exact configuration-space-based branch-and-bound algorithm
(with leaf removal \cite{bauer2001}) 
starts to exhibit a typically exponential running time.

\begin{figure}[htb]
   \includegraphics[width=0.47\textwidth]{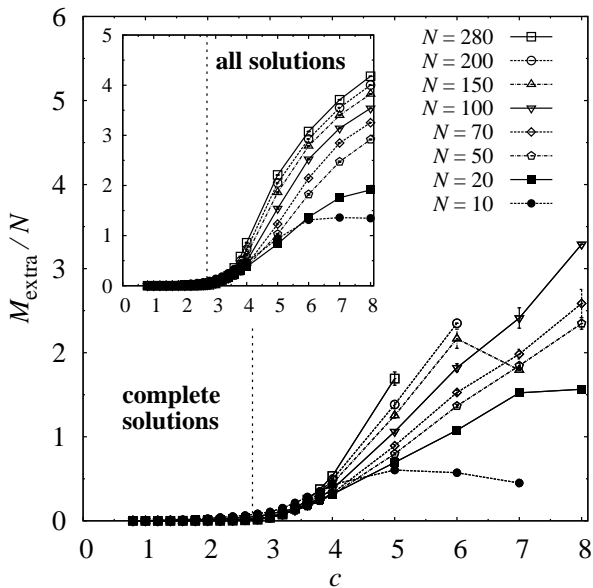}
   \caption{Average number of extra constraints $M_{\text{extra}}$ per
   node for CP algorithm as a function of $c$ only for complete
   solutions and for all
   solutions (inset). The vertical line denotes $c=e$,
the other lines are guides to the eyes only.\label{Extra_constr}}
 \end{figure}

An indicator for the running time  of the SX+CP algorithm is 
the average number of extra
constraints $M_{\text{extra}}$ that were added to the LP
resulting from CPs (\ref{constr}) to obtain complete solutions.
Fig.\ \ref{Extra_constr} shows  $M_{\text{extra}}/N$
as a function of connectivity $c$. Clearly, an increase close
to $c=e$ is visible.  Note that $M_{\text{extra}}/N$ increases
also for all realizations (see inset), but in this case this
is less informative, since the algorithm is stopped if for some
time no new constraint could be added.

\begin{figure}[htb]
   \includegraphics[width=0.47\textwidth]{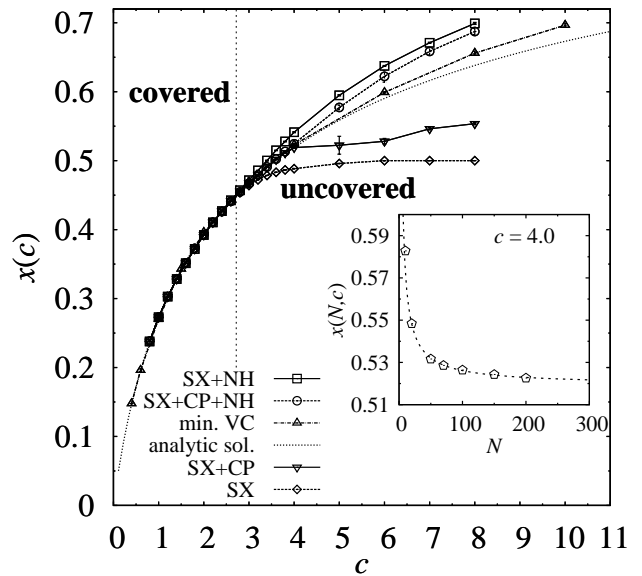}
   \caption{Phase diagram for the fraction of covered vertices $x$.
   Minimum VC found with exact Branch-and-Bound algorithm/analytics
   \citep{cover2000} as well as upper and lower bounds obtained
from the different algorithms. 
   Inset: Finite-size scaling for SX+CP and
   $c=4.0$ The vertical line denotes $c=e$,
dotted line in the inset a result from the fit ($0.52+0.40N^{-0.88}$)
and the other lines are guides to the eyes only.\label{Phasediagr}}
 \end{figure}

Finally, Fig.\ \ref{Phasediagr} shows the phase diagram for the VC problem. In
addition to the exact minimum VC and the analytical solution
\citep{cover2000}, simulation results for the SX and the SX+CP
algorithm (lower bounds) and a combination of SX/SX+CP with NH
(upper bounds) are included. These bounds were obtained, respectively
by averaging the ``cover size'' $x$ for 
different system sizes $N$ and different connectivities
$c$, yielding $x(N,c)$. We extrapolated to infinite system
sizes $x(c) = \lim_{N
\rightarrow \infty} x(N,c)$ via fitting the data to functions
$x(N,c) = x(c) + a \; N^{-b}$ (see inset of figure \ref{Phasediagr})
or $x(N,c) = [x_c(c) + a \; N^{-b}] \cdotp [1+ f \; N^{-g}]$. The
latter was only used for the SX+CP/SX+CP+NH approaches for $c \geq 5$,
where apparently stronger finite-size corrections occur. We found, e.g.,
for SX+CP at $c=4$ a value $b=0.88(9)$ which is compatible with the scaling
of the exact value obtained by branch-and-bound \cite{cover_ccp2005}.

The SX algorithm alone only yields results close to the minimum VC up
to $c = e$. Above this
value the critical fraction of covered vertices converges towards 
the trivial solution $x_c
= 0.5$. The SX+CP approach results in a better lower bound and
deviates only for an average degree $c \geq 5$ visibly from the true minimum
cover sizes. Comparing with the previous results, this means
for $c\gtrsim e$, this approach typically does not yield the correct
solution but comes very close to it. The upper bounds also start
to deviate significantly for $c>e$, but stay rather close.
In general, one sees again the importance of the critical line $c=e$: 
For smaller connectivities $c$ all bounds seem to agree but beyond it
they start to diverge, which is in contrast to previous analytical
bounds \cite{phase-transitions2005}, 
which do not match the correct result for all connectivities $c>0$.

\paragraph*{Conclusion/Outlook}
We studied the vertex-cover problem for Erd\H{o}s-Renyi random graphs 
with a linear programming/cutting plane
algorithm. The algorithm shows a clear ``easy--hard'' signature
close to the connectivity $c=e$. This measn that 
this point denotes a phase transition not only
for configuration-space-based quantities and algorithms 
\cite{cover2000,bauer2001} but
also for the LP/CP approach which operates outside the space 
of feasible solutions. Thus, the typical hardness
of VC is really an intrinsic property of the problem and not
bounded to specific algorithms. This finding may be related
 to the fact that also in the worst-case, all algorithms
known for NP-hard problems exhibit an exponential running time,
i.e. it could help in order to understand better the P-NP problem.

In principle, the number of loops grows exponentially with the number
of nodes. Hence, one could imagine that even within our SX+CP approach,
by exhausting the set of these loop constraints, 
one arrives at an (exponentially slow)
but \emph{complete} algorithm. Nevertheless,
there are graphs, where our constraints are clearly  insufficient: 
A simple example is a complete
graph of size $N=4$, i.e., where each node is connected to any other
node. The result of SX will be to set all $x_i=0.5$, i.e., $x=2$. By
adding constraints for the four possible loops of odd length, the result
will be $x_i=2/3$ for all nodes, i.e., $x=8/3$, while the correct
minimum-cover size is $x=N-1=3$.

One could improve the algorithm in principle by adding
other types of constraints, e.g., general \emph{Gomory-Chv\'atal cuts}
\cite{cook1998}.
Alternatively, one could consider small subgraphs $G'=(V',E')$, $V'\subset V$ 
and $E'\subset E$, solve them by an exact algorithm yielding the cardinality
$X'$ of the minimum cover. Then one could add the constraint $\sum_{i\in V'}
x_i \ge X'$. We have performed some preliminary experiments with these
types of cuts, but observed only marginal improvements
so far, i.e., the overall behavior with the transition close to $c=e$
was preserved.

In practice often a combination of the branching, i.e. configuration-space
based, and cutting, i.e., LP-based approaches, are used. 
Here, branching sets in when all available cutting planes are
exhausted. Thus, it would be very interesting to see how this 
combination of approaches performs
on VC for ER graphs.

Furthermore, it would be of high interest to analytically analyze
the cutting plane approach, to see whether one can understand why
it performs so well for $c<e$. This would lead to a better understanding
of the roots of computational hardness and
 could also lead to refined bounds on the minimum cover sizes, providing
techniques applicable to a vast range of problems.

Finally, it could be worthwhile to extend the present study to
other graph ensembles. One could test whether at the same point
replica symmetry breaking occurs,
hierarchical clustering of solution space can be found, 
and the problem becomes
hard for configuration-space-based branch-and-bound approaches as well as for
the LP-based cutting-plane algorithms.

\begin{acknowledgments}
We thank O. Melchert for critically reading the manuscript and
 M.-T. H\"utt, W. Krauth, 
M. M\'ezard, U. Nowak, D. Sherrington and M. Weigt for stimulating discussions.
The simulations were performed at the HERO cluster of the University
of Oldenburg funded by the DFG (INST 184/108-1 FUGG) and the 
minstry of Science and Culture (MWK) of the Lower Saxony State.
\end{acknowledgments}

\bibliography{alex_refs}

\begin{thebibliography}{25}
\expandafter\ifx\csname natexlab\endcsname\relax\def\natexlab#1{#1}\fi
\expandafter\ifx\csname bibnamefont\endcsname\relax
  \def\bibnamefont#1{#1}\fi
\expandafter\ifx\csname bibfnamefont\endcsname\relax
  \def\bibfnamefont#1{#1}\fi
\expandafter\ifx\csname citenamefont\endcsname\relax
  \def\citenamefont#1{#1}\fi
\expandafter\ifx\csname url\endcsname\relax
  \def\url#1{\texttt{#1}}\fi
\expandafter\ifx\csname urlprefix\endcsname\relax\def\urlprefix{URL }\fi
\providecommand{\bibinfo}[2]{#2}
\providecommand{\eprint}[2][]{\url{#2}}

\bibitem[{\citenamefont{Garey and Johnson}(1979)}]{garey1979}
\bibinfo{author}{\bibfnamefont{M.~R.} \bibnamefont{Garey}} \bibnamefont{and}
  \bibinfo{author}{\bibfnamefont{D.~S.} \bibnamefont{Johnson}},
  \emph{\bibinfo{title}{Computers and intractability}}
  (\bibinfo{publisher}{W.H. Freemann}, \bibinfo{address}{San Francisco},
  \bibinfo{year}{1979}).

\bibitem[{\citenamefont{Papadimitriou and Steiglitz}(1998)}]{papadimitriou1998}
\bibinfo{author}{\bibfnamefont{C.}~\bibnamefont{Papadimitriou}}
  \bibnamefont{and}
  \bibinfo{author}{\bibfnamefont{K.}~\bibnamefont{Steiglitz}},
  \emph{\bibinfo{title}{Combinatorial Optimization -- Algorithms and
  Complexity}} (\bibinfo{publisher}{Dover Publications Inc.},
  \bibinfo{address}{Mineola, NY}, \bibinfo{year}{1998}).

\bibitem[{\citenamefont{Hartmann and Weigt}(2005)}]{phase-transitions2005}
\bibinfo{author}{\bibfnamefont{A.~K.} \bibnamefont{Hartmann}} \bibnamefont{and}
  \bibinfo{author}{\bibfnamefont{M.}~\bibnamefont{Weigt}},
  \emph{\bibinfo{title}{Phase Transitions in Combinatorial Optimization
  Problems}} (\bibinfo{publisher}{Wiley-VCH}, \bibinfo{address}{Weinheim},
  \bibinfo{year}{2005}).

\bibitem[{\citenamefont{M\'ezard and Montanari}(2009)}]{mezard2009}
\bibinfo{author}{\bibfnamefont{M.}~\bibnamefont{M\'ezard}} \bibnamefont{and}
  \bibinfo{author}{\bibfnamefont{A.}~\bibnamefont{Montanari}},
  \emph{\bibinfo{title}{Information, Physics and Computation}}
  (\bibinfo{publisher}{Oxford University Press}, \bibinfo{address}{Oxford},
  \bibinfo{year}{2009}).

\bibitem[{\citenamefont{Moore and Mertens}(2011)}]{moore2011}
\bibinfo{author}{\bibfnamefont{C.}~\bibnamefont{Moore}} \bibnamefont{and}
  \bibinfo{author}{\bibfnamefont{S.}~\bibnamefont{Mertens}},
  \emph{\bibinfo{title}{The Nature of Computation}} (\bibinfo{publisher}{Oxford
  University Press}, \bibinfo{address}{Oxford}, \bibinfo{year}{2011}).

\bibitem[{\citenamefont{Kirkpatrick and Selman}(1994)}]{kirkpatrick1994}
\bibinfo{author}{\bibfnamefont{S.}~\bibnamefont{Kirkpatrick}} \bibnamefont{and}
  \bibinfo{author}{\bibfnamefont{B.}~\bibnamefont{Selman}},
  \bibinfo{journal}{Science} \textbf{\bibinfo{volume}{264}},
  \bibinfo{pages}{1297} (\bibinfo{year}{1994}).

\bibitem[{\citenamefont{Gent and Walsh}(1996)}]{gent1996}
\bibinfo{author}{\bibfnamefont{I.~P.} \bibnamefont{Gent}} \bibnamefont{and}
  \bibinfo{author}{\bibfnamefont{T.}~\bibnamefont{Walsh}}, in
  \emph{\bibinfo{booktitle}{Proceedings of 12th European Conference on
  Artificial Intelligence. ECAI '96}} (\bibinfo{publisher}{Wiley},
  \bibinfo{address}{Chichester}, \bibinfo{year}{1996}), p.
  \bibinfo{pages}{170}.

\bibitem[{\citenamefont{Weigt and Hartmann}(2000)}]{cover2000}
\bibinfo{author}{\bibfnamefont{M.}~\bibnamefont{Weigt}} \bibnamefont{and}
  \bibinfo{author}{\bibfnamefont{A.~K.} \bibnamefont{Hartmann}},
  \bibinfo{journal}{Phys. Rev. Lett.} \textbf{\bibinfo{volume}{84}},
  \bibinfo{pages}{6118} (\bibinfo{year}{2000}).

\bibitem[{\citenamefont{Davis et~al.}(1962)\citenamefont{Davis, Logemann, and
  Loveland}}]{davis1962}
\bibinfo{author}{\bibfnamefont{M.}~\bibnamefont{Davis}},
  \bibinfo{author}{\bibfnamefont{G.}~\bibnamefont{Logemann}}, \bibnamefont{and}
  \bibinfo{author}{\bibfnamefont{D.}~\bibnamefont{Loveland}},
  \bibinfo{journal}{Commun. ACM} \textbf{\bibinfo{volume}{5}},
  \bibinfo{pages}{394} (\bibinfo{year}{1962}), ISSN \bibinfo{issn}{0001-0782},
  \urlprefix\url{http://doi.acm.org/10.1145/368273.368557}.

\bibitem[{\citenamefont{Sedgewick}(1990)}]{sedgewick1990}
\bibinfo{author}{\bibfnamefont{R.}~\bibnamefont{Sedgewick}},
  \emph{\bibinfo{title}{Algorithms in C}} (\bibinfo{publisher}{Addison-Wesley},
  \bibinfo{address}{Reading (MA)}, \bibinfo{year}{1990}).

\bibitem[{\citenamefont{Weigt and Hartmann}(2001)}]{cover-time2001}
\bibinfo{author}{\bibfnamefont{M.}~\bibnamefont{Weigt}} \bibnamefont{and}
  \bibinfo{author}{\bibfnamefont{A.~K.} \bibnamefont{Hartmann}},
  \bibinfo{journal}{Phys. Rev. Lett.} \textbf{\bibinfo{volume}{86}},
  \bibinfo{pages}{1658} (\bibinfo{year}{2001}).

\bibitem[{\citenamefont{Cocco and Monasson}(2001)}]{cocco2001}
\bibinfo{author}{\bibfnamefont{S.}~\bibnamefont{Cocco}} \bibnamefont{and}
  \bibinfo{author}{\bibfnamefont{R.}~\bibnamefont{Monasson}},
  \bibinfo{journal}{Phys. Rev. Lett.} \textbf{\bibinfo{volume}{86}},
  \bibinfo{pages}{1654} (\bibinfo{year}{2001}).

\bibitem[{\citenamefont{Papadimitriou}(1991)}]{papadimitriou1991}
\bibinfo{author}{\bibfnamefont{C.~H.} \bibnamefont{Papadimitriou}}, in
  \emph{\bibinfo{booktitle}{Foundations of Computer Science, Annual IEEE
  Symposium on}} (\bibinfo{publisher}{IEEE Computer Society},
  \bibinfo{address}{Los Alamitos, CA, USA}, \bibinfo{year}{1991}),
  vol.~\bibinfo{volume}{0}, pp. \bibinfo{pages}{163--169}, ISBN
  \bibinfo{isbn}{0-8186-2445-0}.

\bibitem[{\citenamefont{Ardelius and Aurell}(2006)}]{ardelius2006}
\bibinfo{author}{\bibfnamefont{J.}~\bibnamefont{Ardelius}} \bibnamefont{and}
  \bibinfo{author}{\bibfnamefont{E.}~\bibnamefont{Aurell}},
  \bibinfo{journal}{Phys. Rev. E} \textbf{\bibinfo{volume}{74}},
  \bibinfo{pages}{037702} (\bibinfo{year}{2006}).

\bibitem[{\citenamefont{M\'ezard et~al.}(2002)\citenamefont{M\'ezard, Parisi,
  and Zecchina}}]{mezard2002}
\bibinfo{author}{\bibfnamefont{M.}~\bibnamefont{M\'ezard}},
  \bibinfo{author}{\bibfnamefont{G.}~\bibnamefont{Parisi}}, \bibnamefont{and}
  \bibinfo{author}{\bibfnamefont{R.}~\bibnamefont{Zecchina}},
  \bibinfo{journal}{Science} \textbf{\bibinfo{volume}{297}},
  \bibinfo{pages}{812} (\bibinfo{year}{2002}).

\bibitem[{\citenamefont{M\'ezard and Zecchina}(2002)}]{mezard2002b}
\bibinfo{author}{\bibfnamefont{M.}~\bibnamefont{M\'ezard}} \bibnamefont{and}
  \bibinfo{author}{\bibfnamefont{R.}~\bibnamefont{Zecchina}},
  \bibinfo{journal}{Phys. Rev. E} \textbf{\bibinfo{volume}{66}},
  \bibinfo{pages}{056126} (\bibinfo{year}{2002}).

\bibitem[{\citenamefont{M\'ezard et~al.}(1987)\citenamefont{M\'ezard, Parisi,
  and Virasoro}}]{mezard1987}
\bibinfo{author}{\bibfnamefont{M.}~\bibnamefont{M\'ezard}},
  \bibinfo{author}{\bibfnamefont{G.}~\bibnamefont{Parisi}}, \bibnamefont{and}
  \bibinfo{author}{\bibfnamefont{M.}~\bibnamefont{Virasoro}},
  \emph{\bibinfo{title}{Spin glass theory and beyond}}
  (\bibinfo{publisher}{World Scientific}, \bibinfo{address}{Singapore},
  \bibinfo{year}{1987}).

\bibitem[{\citenamefont{Cook et~al.}(1998)\citenamefont{Cook, Cunningham,
  Pulleyblank, and Schriever}}]{cook1998}
\bibinfo{author}{\bibfnamefont{W.~J.} \bibnamefont{Cook}},
  \bibinfo{author}{\bibfnamefont{W.~H.} \bibnamefont{Cunningham}},
  \bibinfo{author}{\bibfnamefont{W.~R.} \bibnamefont{Pulleyblank}},
  \bibnamefont{and}
  \bibinfo{author}{\bibfnamefont{A.}~\bibnamefont{Schriever}},
  \emph{\bibinfo{title}{Combinatorial Optimization}}
  (\bibinfo{publisher}{Wiley}, \bibinfo{address}{New York},
  \bibinfo{year}{1998}).

\bibitem[{\citenamefont{Erd\H{o}s and R\'enyi}(1960)}]{erdoes1960}
\bibinfo{author}{\bibfnamefont{P.}~\bibnamefont{Erd\H{o}s}} \bibnamefont{and}
  \bibinfo{author}{\bibfnamefont{A.}~\bibnamefont{R\'enyi}},
  \bibinfo{journal}{Publ. Math. Inst. Hungar. Acad. Sci.}
  \textbf{\bibinfo{volume}{5}}, \bibinfo{pages}{17} (\bibinfo{year}{1960}).

\bibitem[{\citenamefont{Barthel and Hartmann}(2004)}]{vccluster2004}
\bibinfo{author}{\bibfnamefont{W.}~\bibnamefont{Barthel}} \bibnamefont{and}
  \bibinfo{author}{\bibfnamefont{A.~K.} \bibnamefont{Hartmann}},
  \bibinfo{journal}{Phys. Rev. E} \textbf{\bibinfo{volume}{70}},
  \bibinfo{pages}{066120} (\bibinfo{year}{2004}).

\bibitem[{\citenamefont{Bauer and Golinelli}(2001)}]{bauer2001}
\bibinfo{author}{\bibfnamefont{M.}~\bibnamefont{Bauer}} \bibnamefont{and}
  \bibinfo{author}{\bibfnamefont{O.}~\bibnamefont{Golinelli}},
  \bibinfo{journal}{Eur. Phys. J. B} \textbf{\bibinfo{volume}{24}},
  \bibinfo{pages}{339} (\bibinfo{year}{2001}).

\bibitem[{\citenamefont{Dantzig}(1948)}]{dantzig1948}
\bibinfo{author}{\bibfnamefont{G.}~\bibnamefont{Dantzig}},
  \bibinfo{journal}{Bull. Amer. Math. Soc.} \textbf{\bibinfo{volume}{54}},
  \bibinfo{pages}{1074} (\bibinfo{year}{1948}).

\bibitem[{\citenamefont{Berkelaar et~al.}(2010)\citenamefont{Berkelaar,
  Eikland, and Notebaert}}]{lp_solve}
\bibinfo{author}{\bibfnamefont{M.}~\bibnamefont{Berkelaar}},
  \bibinfo{author}{\bibfnamefont{K.}~\bibnamefont{Eikland}}, \bibnamefont{and}
  \bibinfo{author}{\bibfnamefont{P.}~\bibnamefont{Notebaert}},
  \emph{\bibinfo{title}{Open-source program to solve linear programming
  problems: lp\_solve}},
  \bibinfo{howpublished}{\url{http://lpsolve.sourceforge.net/5.5/}}
  (\bibinfo{year}{2010}), \bibinfo{note}{online; Licence: GNU LGPL (Lesser
  General Public Licence)}.

\bibitem[{\citenamefont{Bland}(1977)}]{bland1977}
\bibinfo{author}{\bibfnamefont{R.}~\bibnamefont{Bland}},
  \bibinfo{journal}{Mathematics of Operations Research}
  \textbf{\bibinfo{volume}{2}} (\bibinfo{year}{1977}).

\bibitem[{\citenamefont{Hartmann et~al.}(2005)\citenamefont{Hartmann, Barthel,
  and Weigt}}]{cover_ccp2005}
\bibinfo{author}{\bibfnamefont{A.~K.} \bibnamefont{Hartmann}},
  \bibinfo{author}{\bibfnamefont{W.}~\bibnamefont{Barthel}}, \bibnamefont{and}
  \bibinfo{author}{\bibfnamefont{M.}~\bibnamefont{Weigt}},
  \bibinfo{journal}{Comp. Phys. Comm.} \textbf{\bibinfo{volume}{169}},
  \bibinfo{pages}{234} (\bibinfo{year}{2005}).

\end{thebibliography}

\end{document}